%% file: main.tex
\documentclass[10pt,conference]{IEEEtran}
\usepackage[letterpaper, top=0.747in, bottom=1.1in, left=0.66in, right=0.66in]{geometry}

\usepackage{graphicx}
\usepackage{balance}
\usepackage{comment}
\usepackage{acro}
\usepackage[acronym]{glossaries}
\usepackage[textfont=md]{subcaption}
\usepackage{xcolor}

\input{acronyms/acronyms}

\author{\IEEEauthorblockN{Seyed Keyarash Ghiasi\IEEEauthorrefmark{1},
Marco Kaldenbach\IEEEauthorrefmark{2}, and Marco Zuniga\IEEEauthorrefmark{3} 
}
\IEEEauthorblockA{Delft University of Technology, The Netherlands\\
Email: \IEEEauthorrefmark{1}s.k.ghiasi@tudelft.nl,
\IEEEauthorrefmark{2}marcokaldenbach@hotmail.com,
\IEEEauthorrefmark{3}m.a.zunigazamalloa@tudelft.nl
\vspace{-10pt}
}
}

\usepackage{cite}
\usepackage{amsmath,amssymb,amsfonts}
\usepackage{algorithmic}
\usepackage{graphicx}
\usepackage{textcomp}
\usepackage{xcolor}
\usepackage{graphicx}
\usepackage{balance}
\usepackage{comment}
\usepackage{acro}
\usepackage[acronym]{glossaries}
\usepackage[textfont=md]{subcaption}
\usepackage{xcolor}
\usepackage{hyperref}

\newcommand{\monitormodel}[0]{\text{LG LM215WF4-TLE7}\,}

\newcommand{\subsectionbeforecut}[0]{\vspace{-5pt}}
\newcommand{\powerreductionpercent}[0]{80\%\,}
\newcommand{\papername}[0]{\text{PassiveCam}\,}
\newcommand{\circled}[1]{\raisebox{.6pt}{\textcircled{\raisebox{-.9pt} {#1}}}}

\def\BibTeX{{\rm B\kern-.05em{\sc i\kern-.025em b}\kern-.08em
    T\kern-.1667em\lower.7ex\hbox{E}\kern-.125emX}}

\begin{document}

\title{Passive Screen-to-Camera Communication\vspace{-10pt}}

\maketitle

\input{chapters/0.Abstract}

%
%

%

\input{chapters/1.Introduction}
\input{chapters/3.Encoding}
\input{chapters/4.Decoding}

\input{chapters/5.Evaluation}
\input{chapters/2.Background}
\input{chapters/6.Conclusion}
\input{chapters/8.Acknowledgement}

\bibliographystyle{IEEEtran}

\appendix
\input{chapters/7.appendix}
\end{document}

%% file: acronyms/acronyms.tex
\DeclareAcronym{led}{
  short=LED,
  long=Light-Emitting Diode,
}

\DeclareAcronym{psr}{
    short=PSR,
    long=Packet Success Rate, 
}

\DeclareAcronym{soa}{
    short=SoA,
    long=state-of-the-art, 
}

\DeclareAcronym{oled}{
  short=OLED,
  long=Organic Light-Emitting Diode,
}

\DeclareAcronym{lcd}{
  short=LCD,
  long=Liquid Crystal Display,
}

\DeclareAcronym{tft}{
  short=TFTs,
  long=Thin Film Transistors,
}

\DeclareAcronym{lc}{
  short=LC,
  long=Liquid Crystal,
}

\DeclareAcronym{fps}{
  short=FPS,
  long=Frames Per Second,
}

\DeclareAcronym{vlc}{
  short=VLC,
  long=Visible Light Communication,
}

\DeclareAcronym{bgr}{
  short=BGR,
  long=Blue Green Red,
}

\DeclareAcronym{cdf}{
  short=CDF,
  long=Cumulative Distribution Function,
}

\DeclareAcronym{fhd}{
  short=FHD,
  long=Full High Definition,
}

\DeclareAcronym{ndk}{
  short=NDK,
  long=Native Development Kit,
}

\DeclareAcronym{sdk}{
  short=SDK,
  long=Software Development Kit,
}

\DeclareAcronym{jni}{
  short=JNI,
  long=Java Native Interface,
}

\DeclareAcronym{jvm}{
  short=JVM,
  long=Java Virtual Machine,
}

\DeclareAcronym{dnn}{
  short=DNN,
  long=Deep Neural Network,
}

\DeclareAcronym{qr}{
  short=QR,
  long=Quick Response,
}

\DeclareAcronym{api}{
  short=API,
  long=Application Programming Interface,
}

\DeclareAcronym{orb}{
  short=ORB,
  long=Oriented FAST and Rotated BRIEF,
}

\DeclareAcronym{rs}{
  short=RS,
  long=Reed-Solomon,
}
\DeclareAcronym{glcm}{
    short=GLCM,
    long=Gray Level Co-occurrence Matrix
}

\DeclareAcronym{cpu}{
    short=CPU,
    long=Central Processing Unit
}
\DeclareAcronym{gpu}{
    short=GPU,
    long=Graphics Processing Unit
}
\DeclareAcronym{vram}{
    short=VRAM,
    long=Video Random-Access Memory
}

\DeclareAcronym{ber}{
    short=BER,
    long=Bit Error Rate
}

\DeclareAcronym{ram}{
    short=RAM,
    long=Random-Access Memory
}
\DeclareAcronym{roi}{
    short=ROI,
    long=Region of Interest
}
\DeclareAcronym{cuda}{
    short=CUDA,
    long=Compute Unified Device Architecture
}
\DeclareAcronym{snr}{
    short=SNR,
    long=Signal-to-Noise-Ratio
}

%% file: chapters/0.Abstract.tex
\begin{abstract}
A recent technology known as \textit{transparent screens} is transforming windows into displays. These smart windows are present in buses, airports and offices. They can remain transparent, as a normal window, or display relevant information that overlays their panoramic views. 
In this paper, we propose transforming these windows not only into \textit{screens} but also into \textit{wireless transmitters}. To achieve this goal, we build upon the research area of screen-to-camera communication. In this area, videos 
are modified in a way that smartphone cameras can decode data out of them, while this data remains invisible to the viewers. 
A person sees a normal video, but the camera sees the video plus additional information.
In this communication method, one of the biggest disadvantages is
the traditional screens' power consumption, more than 80\% of which is used to generate light. To solve this, we employ novel transparent screens relying on ambient light to display pictures, hence eliminating the power source. However, this comes at the cost of a lower image quality, since they use \textit{variable} and out-of-control environment light, instead of generating a 
\textit{constant and strong} light by LED panels.
Our work--dubbed \papername--overcomes the challenge of creating the first screen-to-camera communication link using passive displays. This paper presents two main contributions. First, we analyze and modify existing screens and encoding methods to embed information reliably in ambient light. Second, we develop an Android App that optimizes the decoding process obtaining a real-time performance.
Our evaluation, which considers a musical application, shows a \ac{psr} of close to 90\%. In addition, our real-time application achieves response times of 530\,ms and 1071\,ms when the camera is static and when it is hand-held, respectively.  

\end{abstract}


%% file: chapters/1.Introduction.tex
\vspace{-8pt}
\section{Introduction}
\label{sec:introduction}
In today's technology-driven world, novel communication methods are becoming prevalent. An emerging example is screen-to-camera communication, where data is embedded into videos in a way that is invisible to viewers, but not to cameras. This opens up opportunities to increase communication capacity, since these screens are pervasive and operate on visible light, instead of the overly-congested and expensive radio spectrum.
Motivated by these advantages, there are several studies in screen-to-camera communications \cite{aircode,chromacode,inframe,inframepp,texturecode}, whose focus is on increasing the \ac{psr} without affecting the viewer's experience. 

The research community has achieved large feats in screen-to-camera communication, but the main downside is the power cost. As an example, a regular LCD monitor uses at least 26 Watts\footnote{Based on measurements on a \monitormodel}. The culprit is its back light, which consumes more than \powerreductionpercent of the total power. A new generation of transparent screens are removing that power cost by using ambient light instead of LEDs. The image quality is not as sharp as in traditional screens but they consume much less energy, which is a major advancement in designing novel screens with a low ecological footprint.

\begin{figure}[t]
    \centering        \includegraphics[width=0.5\columnwidth]{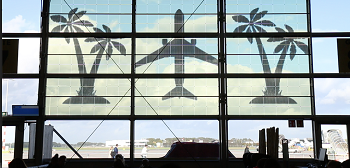}
        \caption{An application of passive displays: a video is shown in an airport using ambient light. The display remains transparent and does not block the outside's view. Image from \cite{Videowindow_website}.} 
    \label{fig:airport_application}
    \vspace{-15pt}
\end{figure}

Transparent screens are already being deployed in buses, airports and offices despite being a novel technology. They are transforming the facades of different environments to transmit relevant information on demand. For example, smart windows facing the runway of airports display advertisements to the passengers, as shown in Fig. \ref{fig:airport_application}.
However, displaying plain pictures and videos on these monitors is not exploiting their potential. In other words, what if the transparent screen transmitted more information? For instance, while showing an enticing holiday destination some users may want more information on discounted hotels or restaurants in that specific location. To achieve that, a user could point her smartphone's camera towards the video and obtain that \textit{embedded and invisible} information.  

Inspired by advances in screen-to-camera communication, we present \papername, a system that enables the first wireless link between transparent screens and smartphones. While being advantageous in terms of power consumption, designing this new link introduces two main challenges. First, compared to the original back light of a display, ambient light is weaker and variable, resulting in lower image quality, which in turn makes the communication link more error-prone. Second, most \ac{soa} works do not consider proper human-computer-interaction designs. Some studies do not evaluate the effect of a user's camera movements, evaluating their work only with their camera on a tripod; other studies do not process the videos in real time; and a few systems rely on visual markers, either static or dynamic, to delineate the borders of their transmitting area, affecting the system's aesthetics.

To overcome the limitations mentioned above, in this paper, we propose a novel communication system with a passive screen acting as transmitter and a smartphone (App) acting as a receiver. 
In general, \papername provides the following contributions: 

\noindent\textbf{\textit{1) A passive screen transmitter.}} Passive screens are not off-the-shelf components yet, but we test two transparent screens: one obtained from a specialized vendor~\cite{Videowindow_website} and the other created by us. We describe how to remove the back-light of normal displays so the interested researchers can create their own passive screen. Our modified screen reduces the power consumption by \powerreductionpercent. Regarding the modulation process, we build on top of the \ac{soa}, however, we improve their encoding with three steps: Gaussian kernels, step encoding, and an optimization of texture analysis.



\noindent\textbf{\textit{2) A reliable real-time receiver.}} Our receiver, running on a custom Android App, tackles two important points to facilitate a seamless user experience. First, unlike many prior studies where a static phone on a tripod is used, we implement a simple hand motion filter so that users can hold the phone. Second, to attain a real time response, we propose a new method to detect the \ac{roi} and discard invalid frames. 

\noindent\textbf{\textit{3) A new application:}} As far as we know, the use of passive screens for communication (commercial or modified) has not been explored before. We evaluate the link's reliability and delay with controlled experiments, but we also propose a novel application. Together with personnel of a company, we target the following scenario. Considering the massive windows (passive screens) at airports, multiple types of videos could be displayed (music, tourism, news, etc) and users may want to listen to one of those streams. To do so, they could point their phone toward the desired video, and our passive link would decode the video ID and the exact time of the frame to provide its synchronized audio. A prototype of such application is shown in two videos: one in \href{https://youtu.be/S_hxk3TQCmI}{black-and-white}\cite{video_black_white} and the other in \href{https://youtu.be/7FoaacEiG60}{color}\cite{video_color}. Overall, \papername has an offline code success rate of close to 90\%. Besides, our real-time Android application acquires synchronization within 530\,ms when mounted on a tripod and 1071\,ms when hand-held.



%% file: chapters/3.Encoding.tex
\vspace{-5pt}
\section{A Passive transmitter}
\label{sec:transmitter}
\begin{figure}[t]
    \centering        \includegraphics[width=0.9\columnwidth]{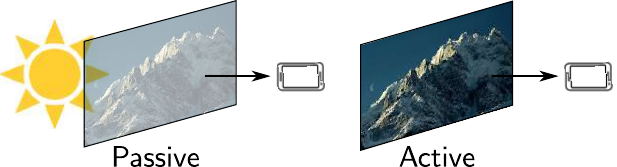}
        \caption{In passive displays, the backlight is provided by sunlight. In standard (active) displays, the backlight is provided by \ac{led}s. If data is embedded in the frames, a smartphone can decode information with screen-to-camera communication.} 
    \label{fig:block_diagram}
    \vspace{-10pt}
\end{figure}

In order for a display to qualify as ``passive'', it requires using ambient light, in contrast to \textit{normal} \ac{lcd}s emitting light by themselves. Fig. \ref{fig:block_diagram} shows the difference between active and passive screens. Transparent screens are not widely sold in the consumer market except by few select companies. Therefore, we will first describe how a passive screen can be obtained from a normal screen, and then, we will explain the modulation process. 
\subsectionbeforecut
\subsection{Passive displays}
\begin{figure}[t]
    \centering
        \includegraphics[width=0.48\columnwidth]{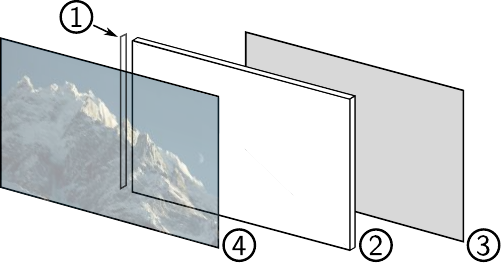}
        \caption{Layers and components inside an off-the-shelf monitor: \circled{1} an \ac{led} or  lamp, \circled{2} a light dispersion layer, \circled{3} metallic reflective layer, \circled{4} a transparent \ac{lcd} glass containing pixels} 
    \label{fig:monitor_exploded}
    \vspace{-17pt}
\end{figure}

To understand how a passive screen can be made, first, we need to explain the structure of a typical display. \ac{lcd}s are made of several components as shown in Fig. \ref{fig:monitor_exploded}. \circled{1} is an array of \ac{led}s or a fluorescent lamp emitting the monitor's backlight. Layer \circled{2} is an acrylic layer dispersing the light uniformly across the monitor. The dispersive layer scatters light to both the back and front, but the light that goes into the back side of the monitor is wasted. 
Therefore, a reflective layer (\circled{3}) is used to recover this light and send it towards the viewer. At the end, layer \circled{4} is a transparent glass layer containing pixels and displays the image to the user.

Based on our measurements, the screen's backlight consumes more than 80\% of the total \ac{lcd}'s power. Thus, to make a passive screen, we need to remove all components except layer \circled{4}. It is important to note that the interface that controls the display is only connected to layer \circled{4}, which contains the pixels. Thus, it can still show pictures and videos. 
The process required to disassemble a screen is not complex and can be done with any \ac{lcd}, however, it has to be done carefully to avoid damaging the glass layer. An example of our passive screen, placed over a window, is shown in the videos of \cite{video_black_white} and \cite{video_color}.

Another type of passive display we use is a ``smart window''\cite{Videowindow_website}. 
Smart windows have similar operating principles to the layer \circled{4} of a monitor. Nevertheless, they are optimized to cause \textit{less attenuation} to the light, resulting in more transparency and a \textit{higher contrast}. 
This lower attenuation of light comes at a cost: smart windows only provide black-and-white images. To show colored pictures, normal monitors have color filters on their pixels, which dissipate light excessively. As a result, a smart window is made without these color filters. On the contrary, a modified \ac{lcd} \textit{looks} darker than a smart window, providing in a lower \ac{snr} when used as a transmitter.

Overall, a smart window is optimized to operate as a black-and-white transparent screen, however, they are not easily accessible yet. Therefore, we evaluate both types of displays. Besides, we develop our final application on a modified display to show that a passive link can also be made without a high-end smart window.
\vspace{-5pt}
\subsection{Data embedding}
\label{sec:encoding}
Embedding data into videos has been considered by several works~\cite{inframe,inframepp,aircode,chromacode}, all of which use active screens. In this section, we first present the basic principles of data embedding, and then, we describe how we build upon these \ac{soa}. 

\subsubsection{Basic principles} Data embedding exploits the slow response of human eyes. If a sequence of colored pixels are shown fast enough, the human eyes would perceive the average color. Based on this principle, we present a general encoding algorithm that reduces visual effects to human eyes. 

An algorithm that embeds data takes two main inputs: a video with frames indexed by $f_0, ... f_n$, and the \textit{data} that has to be encoded in those frames. In visual communications, data is represented by \textit{binary frames}.
A binary frame is similar to a \ac{qr} code showing a checkerboard pattern. To take advantage of the \textit{basic principle} above, the data encoding algorithm works as follows. First, every frame in the original video is duplicated. Then,
a modulation value of $\Delta$ is calculated for each pixel. The pixel at location (x,y)--denoted by $f_i(x,y)$--in the first frame is then increased by its respective $\Delta(x,y)$, i.e. $f_i^+(x,y) = f_i(x,y) + \Delta(x,y)$. In the other frame, the modulation values are subtracted, i.e. $f_i^-(x,y) = f_i(x,y) - \Delta(x,y)$. The resulting $f_i^+$ and $f_i^-$ are called complementary frames, and if played fast enough, the viewer's eyes shall average them to see the frame $f_i$, although modulated frames are being displayed.

Regarding the creation of binary frames, the \ac{soa} proposes custom procedures, however, we opt for \ac{qr} codes for two reasons: (1) They are mature and well-developed, thus can be easily integrated into an application, and (2) The implementations are already optimized and tested, hence reducing the overhead of designing a receiver pipeline.  

\subsubsection{Lightness modulation}

The binary frame indicates the location of the pixels that need to be modulated, but it does not state how to derive $\Delta(x,y)$. For this, the \ac{soa} proposes various modulation schemes~\cite{inframe,inframepp,aircode,chromacode}. We build upon schemes that modulate \textit{lightness} instead of \textit{color}, as our smart window only operates in black-and-white. 

Several approaches use color spaces to guide the modulation process. The CIELAB standard divides the color space into two main dimensions: lightness ($L^*$) and color ($a^*, b^*$). In our case, 
to modulate a given pixel, the lightness is changed by $\Delta L^*$. Following the complementary modulation above, if a pixel falls in a white region of the \ac{qr} code, we set the its $\Delta(x,y)$ to $+\Delta L^*$. Otherwise, it is set to $-\Delta L^*$.
Therefore, for each frame, we two complementary modulated frames are generated.
Next, we first describe the building blocks we borrow from the \ac{soa}, ChromaCode~\cite{chromacode}, and after that, we describe our improvements.

\textbf{Building blocks from the \ac{soa}.} ChromaCode's key contribution is to show that the parameter $\Delta L^*$ depends on the \textit{human perception of color} and the \textit{texture} around a pixel. 
The human perception is self explanatory. Regarding the texture, for instance,
changes in lightness are less perceptive in \textit{rough} images (like rocks) than in \textit{soft images} (like clear skies). 
Thus, the same modulation depth $\Delta L^*$ would affect the flicker effect of each pixel differently.

\vspace{2pt}
\noindent\textit{\underline{Step 1: Perception-based derivation of $\Delta L$}.}
To determine a proper change of lightness $\Delta L_{1}^{*}$ for a pixel location (x,y), ChromaCode
employs the CIEDE2000 formula to take human perception into account. 
The complete formula is explained in \cite{chromacode}. Below, we present a re-arranged equation that suits our purpose. Denoting $L_{1}^{*}(x, y)$ as the lightness of pixel $(x, y)$, $k_L$ as a parameter to compensate for different viewing conditions, and $\Delta E_{00}$ as the perceived difference in color between two \textit{complementary} pixels; the modulation depth $\Delta L_{1}^{*}(x, y)$ is given by: 

\vspace{-20pt}
\begin{equation}
\Delta L_{1}^{*}(x, y) = k_L[ 1 + \frac{0.015(L^{*}(x, y) - 50)^2}{\sqrt{20 + (L^{*}(x, y) - 50)^2}} ] \Delta E_{00}
\label{eq:dL1}
\end{equation}
\vspace{-5pt}

In the above equation, $L_{1}^{*}(x, y)$ is given for every pixel and $k_L$ is set to 1. The important part is to determine $\Delta E_{00}$. A large value increases the signal's \ac{snr} but also the flicker effect. A $\Delta E_{00}$ between 1 and 2 results in acceptable flicker levels. 

\vspace{2pt}
\noindent\textit{\underline{Step 2: Texture-based adaptation of $\Delta L$}.}
In the prior step, $\Delta L^*$ considers only the perceived difference in color $\Delta E_{00}$, but not texture.
This step estimates the texture of a pixel area by measuring how abruptly the pixels change. The bigger the changes, the rougher the texture, and the less obvious the flicker effects.

A common approach to calculating texture is
Regional Texture Analysis, which constructs a \ac{glcm} for every pixel. Based on this matrix, one can calculate the contrast $C(x, y)$ in a pixel area.
The construction of the \ac{glcm} matrix and contrast is detailed in the appendix. 
Using this method, ChromaCode\cite{chromacode} further proposes an \textit{average} texture metric $T(x, y)$ around pixel \textit{p(x,y)}:

\vspace{-5pt}
\begin{equation}
T(x, y) = \frac{C(x, y)}{S(x, y)}
\label{eq:texture}
\end{equation}
\vspace{-10pt}

where $C(x, y)$ is the contrast and $S(x, y)$ is the number of pixels in a $n\times n$ area around pixel $(x,y)$. Based on $T(x, y)$, the following scaling parameter $\alpha(x, y)$ is derived to adjust the modulation depth based on texture: 

\vspace{-5pt}
\begin{equation}
\alpha(x, y) = \frac{T(x, y)}{T_{max}} (1 - k) + k,
\label{eq:dL2_scaling_factor}
\end{equation}
\vspace{-15pt}

where $T_{max}$ is the maximum value of $T(x,y)$ in the frame, and $k$ is a parameter with a default value of 0.5. 

In the end, the modulation depth of a pixel $\Delta L_{2}^{*}(x, y)$ is determined by the \textit{color perception} $\Delta L_{1}^{*}$ and \textit{texture} $\alpha(x, y)$.

\vspace{-10pt}
\begin{equation}
\Delta L_{2}^{*}(x, y) = \Delta L_{1}^{*}(x, y) \cdot \alpha(x, y).
\label{eq:dL2}
\end{equation}
\vspace{-15pt}

This value modifies the lightness of each pixel in the video.

\textbf{Improvements over the \ac{soa}.} The \ac{soa} provides a valuable starting point to embed information, but there are a few shortcomings that need to be improved for \papername. 

The \ac{soa} relies on better transmitters, with constant strong light; and better receivers, with cameras having 120 or 240 \ac{fps}. Our passive transmitter has weak variable light, and the phone we use has 60 \ac{fps} (to develop a more inclusive system). These two differences cause more flicker effects due to variable lighting conditions and a slower modulation.
To ameliorate these flicker effects, we propose two improvements.

\vspace{2pt}
\noindent\textit{\underline{Improvement 1: Gaussian kernels.}} Given that the \textit{sharp} borders in the binary frames cause noticeable artefacts, 
we apply a Gaussian kernel over the binary frame to smooth the sharp transitions. This process blurs the borders, making the embedding less noticeable, as surveyed in Section \ref{sec:survey}.

\begin{figure}[t]
    \centering
    
        \centering
        \includegraphics[width=0.55\columnwidth]{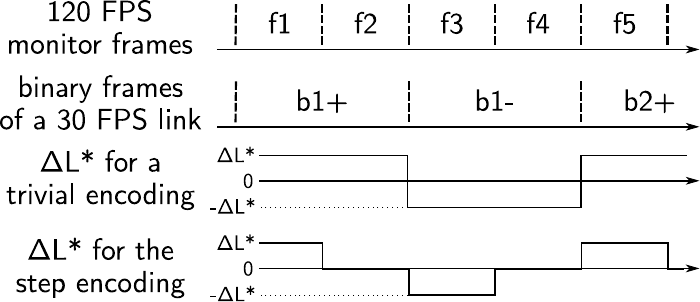}
    \caption{Timeline of step encoding versus standard encoding. 
    }
    \label{fig:step_encoding_vs_trivial}
    \vspace{-15pt}
\end{figure}

\vspace{2pt}
\noindent\textit{\underline{Improvement 2: Step encoding.}} 
In screen-to-camera communication, the frame rate of the receiver (camera) must be the same or higher than the transmitter's (screen).
There are still many smartphones limited to capturing at 30 \ac{fps} in real-time. In this case, the screen must also transmit at a maximum of 30 \ac{fps}. This limitation causes the capacity of most screens with 60 or 120 \ac{fps} to be underutilized. Therefore, we propose a method to use this extra frame rate for the sake of flicker reduction. In the \ac{soa}, a modulated pixel changes from $+\Delta L^{*}$ to $-\Delta L^{*}$, making for an abrupt change of $2\Delta L^{*}$. 
However, if the screen can work at a higher rates than the camera, the encoding can be done in smaller, giving the impression of a smoother transition. 
For instance, let us consider a display with 120 \ac{fps}. If we aim for the transmission of binary frames at 30 \ac{fps}, a trivial configuration is to have two frames with $+\Delta L^{*}$ and two frames with $-\Delta L^{*}$. On the contrary, 
we can have one frame with no modulation, one frame with $+\Delta L^{*}$, another frame without modulation, and the last frame with $-\Delta L^{*}$. In this way, the lightness transition between frames is reduced by half, from $2\Delta L^{*}$ to $\Delta L^{*}$, ameliorating flicker effects. The trivial modulation and and the step encoding are both shown in Fig. \ref{fig:step_encoding_vs_trivial} for this example.

\vspace{2pt}
\noindent\textit{\underline{Improvement 3: Optimized texture analysis.}} Our third improvement is related to optimizing the computational performance. Embedding information on videos requires several signal-processing steps. The most demanding step is the texture analysis done with the \ac{glcm} matrices. This high overhead would not be a problem if data was embedded only once. However, for applications that require periodic encoding, potentially in real-time, it is worth using the optimized method presented in appendix
\ref{app:optimize_constrast}, 
which can reduce the encoding time by a factor of 500. 

%% file: chapters/4.Decoding.tex
\vspace{0pt}
\section{Decoding}
\vspace{-4pt}
\label{sec:decoding}
To decode a video, a smartphone's camera is used. We consider two scenarios: \textit{offline} and \textit{real-time} decoding. The \textit{offline} process first records the video and then recovers data. In contrast, the \textit{real-time} process starts recovering data as soon as the frames are captured. The \textit{real-time} process targets realistic scenarios of a user receiving data from the screen, and the \textit{offline} process aims at evaluating the encoder's performance, regardless of the processing power of a user's phone.

\subsectionbeforecut
\subsection{Offline decoding}
\label{sec:offline_decoding}
For the offline decoding, we mount a smartphone on a tripod, and its camera records the encoded video, as described in Section \ref{sec:encoding}. Next, the files are transferred to a PC for data extraction using the following processing steps:  

\noindent\underline{Step 1) Adjusting the camera's frame rate:}
As stated by the Nyquist law, the sampling rate should be at least twice the highest frequency in the transmitted signal. With complementary modulation, the highest frequency 
is half the modulation's frame rate. Thus, if the camera captures data with at least the same frame rate as the video, Nyquist is satisfied. 
Capturing at a higher rate can be advantageous as there is less chance of frame loss due to camera jitter. However, rolling shutter effects might become present.   



\noindent\underline{Step 2) Adjusting the camera's ISO and shutter speed:}
ISO is the sensitivity of the camera's sensor to light, and the shutter speed is the duration of the sensor's exposure to light. 
In CMOS image sensors, there is a chance of capturing a rolling shutter effect, depending on the shutter speed. Usually, increasing the shutter speed (reducing the exposure) removes this effect. However, it darkens the pictures, resulting in the loss of encoded data. Hence, by increasing the ISO, brightness can be adjusted again. Based on the ambient light, these values can differ. In Section \ref{sec:evaluation}, we do experiments using different levels of ISO and shutter speed to find the best setup.

\noindent\underline{Step 3) Data extraction:}
In Section \ref{sec:encoding}, we mentioned that the encoding algorithm makes complementary frames: if in frame \textit{i} (denoted by $f_i^+$), the pixel (x,y) has a lightness of $L_{i}(x,y)^* \pm \Delta L_{i}(x,y)^*$, the lightness of the (x,y) pixel in its complementary frame (denoted by $f_i^{-}$) will be $L_{i}(x,y)^* \mp \Delta L_{i}(x,y)^*$. Therefore, subtracting $f_i^{+}$ and $f_i^-$ yields a new frame with pixel values of $\pm 2\times \Delta L_{i}(x,y)^*$. This last frame contains the encoded \ac{qr} information. Considering this differential process, we implement the decoder.

Following Step 1, we assume that the camera's frame rate is equal to the transmitter's, thus, each frame is captured once and no frames are missed\footnote{This assumption helps simplify the explanation. In practice, some frames will be missed, but they can be recovered using error correction codes}.  
As a result of the matching rate between the screen and camera\footnote{Later, we generalize the decoding procedure for higher camera frame rates}, the receiver has all the frames and their complements in order, i.e. $\{..., f_{i-1}^+ , f_{i-1}^- , f_{i}^+ , f_{i}^-, ...\}$. 
Then, we subtract all captured frames from the one before. Our goal is to keep the result when a frame is subtracted from its complement. However, an important point is that the receiver cannot know in advance which frames are complementary. Therefore, in every other subtraction, it operates on frame \textit{i} ($f_i^+$) and the complement of the previous (non-matching) frame ($f_{i-1}^-$), yielding a corrupt \ac{qr} code since two different binary frames are subtracted. As we will show in \underline{Step 5}, these corrupt frames can be detected and do not pose any problems besides consuming the decoder's time.

To generalize our assumption about the frame rates, if the camera captures at a higher rate, there will be more instances of the same frame. In that case, subtracting consecutive frames sometimes results in a dark image, i.e. when the captured instances of the same frame are subtracted. 

\noindent\underline{Step 4) \ac{roi} detection:} Since the binary codes may not cover the entire frame, a captured video will likely contain both encoded pixels and other surroundings, including non-modulated regions of the video. Before passing the pixels to a \ac{qr} decoder, the region containing valid data has to be identified. Previous works use visual markers around a screen so that the \ac{roi} can be easily detected, but we use a simpler technique. Going back to \underline{Step 3}, we notice that in subtracted frames, all regions that are not encoded (shifted by $\Delta L^*$) have a lightness of 0, as their pixel values are the same as in their complementary frames. Using this fact to determine our \ac{roi}, we just need to find the smallest region inclusive of all pixels with a non-zero lightness, which is a simple task in image processing.




\noindent\underline{Step 5) \ac{qr} decoding:}
This is the last step to extract binary data from pictures. For offline decoding, we use a \ac{qr} decoder library in Python, and process the \ac{roi}s found in \underline{Step 4}. However, as mentioned in \underline{Step 3}, every other subtracted frame contains corrupt data. This does not pose any problems in offline decoding, since \ac{qr} codes have checksums and corrupt frames can be detected and dropped. However, attempting to decode those frames delays the processing, which becomes problematic in a real-time system, which will be discussed below. 
\subsection{Real-time decoding}
 After validating the encoding with the offline process, we implement the receiver pipeline on a smartphone to build a real-time link. When a person uses the phone in real-time, the receiver faces extra challenges mentioned below.


\subsubsection{Hand movement filtering} 
A user, unlike a tripod, cannot hold a phone completely steady. Therefore, we have to remove the effect of camera movements before decoding. 
To address this dynamic, we implement a simple hand motion filter that uses prominent points in consecutive frames as \textit{anchors}, and calculates a matrix to map these prominent features between frames. By applying the inverse transform to the shifted image, its pixels can align well with their counterpart in the other frame, hence alleviating the hand movement effect.  





\subsubsection{Latency optimization}
As mentioned before, subtracting two non-complementary frames yields undecodable \ac{qr} codes. Following the method in \underline{Step 3}, undecodable codes would appear in every other subtracted frame. If the corrupt data is speculated preemptively, it can be discarded right away, without waiting for the \ac{qr} decoder to do so. Such receiver is called to be \textit{synchronized}. 

To acquire synchronization, we build upon the observation that the correct frames are alternating, therefore, at the beginning, the pipeline makes an attempt to decode all frames, corrupt and healthy, but soon after the first successful decoding, the pipeline sends every other frame to the decoder, and discards the rest. Some factors, such as camera jitter, might result in losing this synchronization over time. Therefore, we should perform periodic checks, and if the decoding error is above a threshold, a synchronization has to be performed again.

 


%% file: chapters/5.Evaluation.tex
\section{Evaluation}
\vspace{-9pt}
\label{sec:evaluation}
\subsection{Core setup}
\begin{figure}[t]
    \centering
    \begin{subfigure}{0.47\columnwidth}
        \centering
        \includegraphics[width=1\columnwidth]{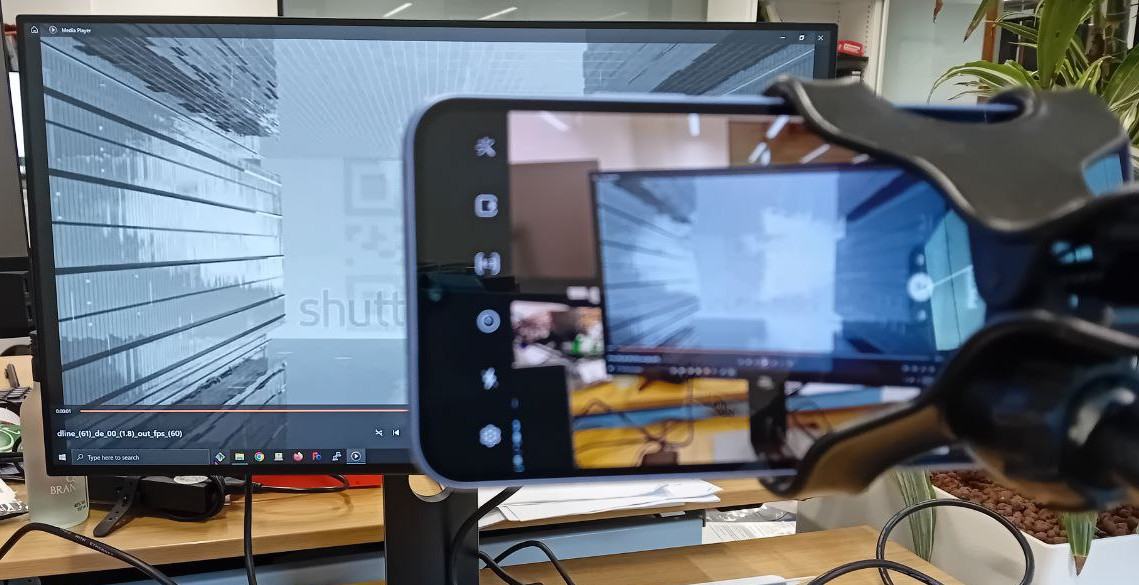}
        \caption{Active monitor}
        \label{fig:setup_active}
    \end{subfigure}
    \begin{subfigure}{0.45\columnwidth}
        \centering
\includegraphics[width=1\columnwidth]{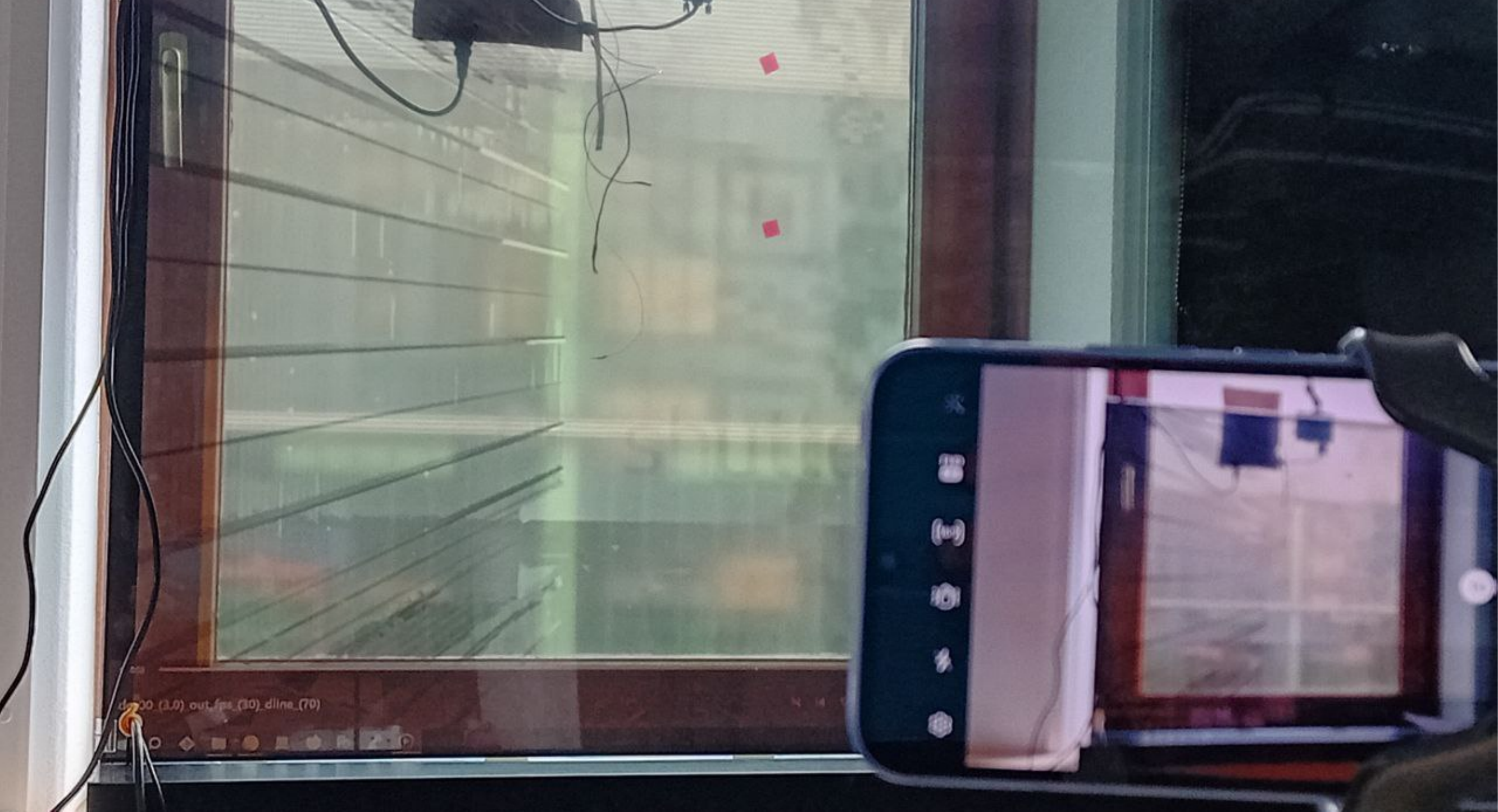}
        \caption{Smart window}
        \label{fig:setup_passive}
    \end{subfigure}
    \vspace{-5pt}
    \caption{Setup of the experiments.}
    \label{fig:experiment_setup}
    \vspace{-15pt}
\end{figure}

In our experiments, we use three screens: The first screen is an active (normal) \ac{lcd} that we use as a baseline to compare with passive displays (Fig. \ref{fig:setup_active}), the second one is a modified passive display obtained 
from an \monitormodel monitor using the procedure of Section \ref{sec:transmitter}, and the third one is a smart window obtained from a manufacturer~\cite{Videowindow_website}. The passive screen and the smart window do not emit any light, therefore, we install them in front of our office's window, as shown in Fig. \ref{fig:setup_passive}. 
On the receiver side, a smartphone camera is placed at various distances up to 1.5\,m from the screen. 
Using this setup, we carry out the experiments described below.


\subsectionbeforecut
\subsection{Performance in cloudy winter days}

The fundamental difference between active and passive screens is the backlight source. During the day, sunlight can range from one thousand Lux on a very cloudy day to 100\,kLux on a clear day. With normal sunlight, our modified passive screen and the smart window work well. However, it is important to assess how different passive screens perform under unfavorable lighting conditions. 

To have a controlled scenario, we use offline decoding during a \textit{cloudy winter day}. In these tests, we encode a single \ac{qr} code in a video with different levels of visibility. We sweep the $\Delta E_{00}$ value in Equation \ref{eq:dL1} from 1.0 to 3.0 in steps of 0.2, yielding different $\Delta L^*$ to embed the \ac{qr} codes. This video is displayed on the three monitors explained above, and the results are discussed next.

\begin{figure}[t]
    \centering
    \begin{subfigure}{0.48\columnwidth}
        \centering
        \includegraphics[width=1\columnwidth]{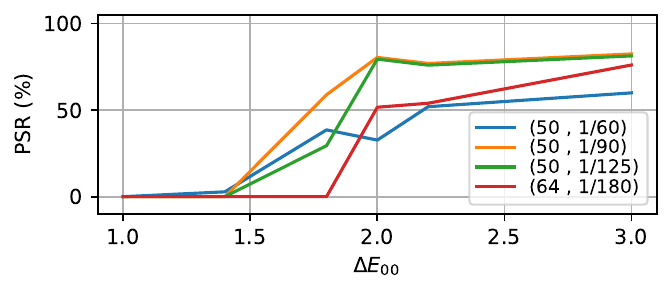}
        \caption{Active monitor at 60 Hz}
        \label{fig:active_offline}
    \end{subfigure}
    \begin{subfigure}{0.48\columnwidth}
        \centering
\includegraphics[width=1\columnwidth]{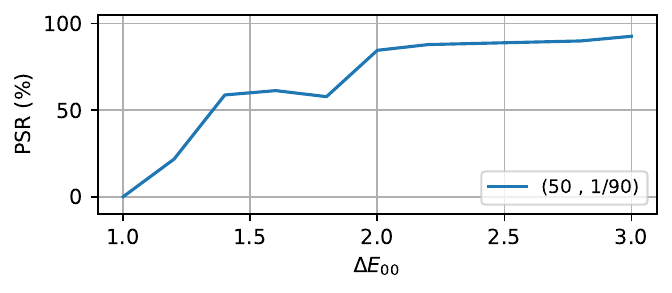}
        \caption{Passive Monitor at 30 Hz}
        \label{fig:passive_offline}
    \end{subfigure}
    \vspace{-5pt}
    \caption{Decoding performance on active and passive monitors.} 
    \label{fig:gaussian}
    \vspace{-15pt}
\end{figure}

\subsubsection{Active display}
First, we use the \textit{active} monitor as a benchmark with a frame rate of 60 \ac{fps}. A camera is installed on a tripod at a distance of 50\,cm from the monitor capturing videos at 60 \ac{fps}. For each transmitted video, we set the camera to work with four different combinations of ISO and shutter speeds. Then, the decoding pipeline mentioned in Section \ref{sec:offline_decoding} is used to calculate the \ac{psr}. Fig. \ref{fig:active_offline} shows the link's reliability (\ac{psr}) as a function of modulation depth  ($\Delta E_{00}$). The four camera settings are denoted by the tuple (ISO, shutter speed), with the shutter speed being in seconds. Among the four combinations, (50,1/90) results in the best \ac{psr}, which is about 80\% for $\Delta E_{00}$ values above 2.0. However, if $\Delta E_{00}$ is above 1.8, the encoding becomes gradually visible to people. The problem is that if we use this lower value of $\Delta E_{00}$, the \ac{psr} drops to about 50\%. Hence, overall, with an active screen at 60 \ac{fps}, we get around 80\% \ac{psr} but causing some flicker.





\subsubsection{Smart window}
For this step, the videos are the same but the smart window is bigger than the active display. To capture the whole monitor, we place the camera at a distance of 1.5\,m. The smart window has a maximum frame rate of 30 \ac{fps}, therefore, we also capture videos at 30 \ac{fps}. The results are shown in Fig. \ref{fig:passive_offline}. In these experiments, only one (ISO, shutter speed) setting worked reliably. The results are better than with the active monitor, around a 90\% \ac{psr} with a $\Delta E_{00}$ above 2.0. This improvement is due to the lower frame rate since it ameliorates the rolling shutter effect.
Overall, despite the low sunlight conditions, the active screen and smart window provide comparable performance. This is thanks to the smart window's emphasis on reducing attenuation but at the cost of working only in black-and-white. 

\subsubsection{Modified screen}
The active display and the smart window could transmit data packets with various \ac{psr}s, as discussed earlier. However, our modified screen could not operate reliably because it attenuates more light than the smart window, as mentioned in Section~\ref{sec:transmitter}. 
This low performance, however, is only for very low lighting conditions, the demo in the videos is done with our modified screen under normal sunlight conditions. The advantage of our modified screen is that it is obtainable from any monitor.


\newcommand{\evalfigurecoeff}[0]{0.24}
\newcommand{\evalcaptionvspace}[0]{-17pt}
\newcommand{\evalfigwidthcoeff}[0]{1}
\newcommand{\totalevalvspace}[0]{-2pt}
\begin{figure*}[t]
    \centering
    \begin{subfigure}{\evalfigurecoeff\textwidth}
    \centering
    \includegraphics[width=\evalfigwidthcoeff  \columnwidth]{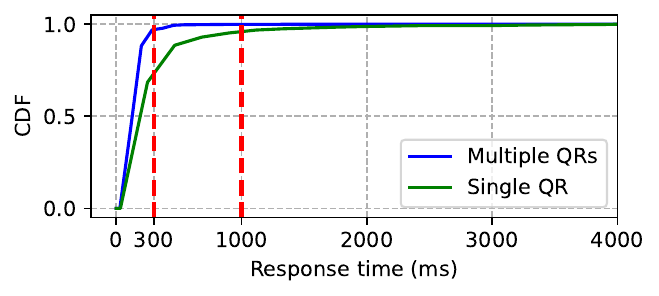}
        \vspace{\evalcaptionvspace}
    \caption{Single QR vs. Multiple QR}
    \label{fig:single_qr_vs_multiple}
\end{subfigure}
    \begin{subfigure}{\evalfigurecoeff\textwidth}
        \centering
\includegraphics[width=\evalfigwidthcoeff\columnwidth]{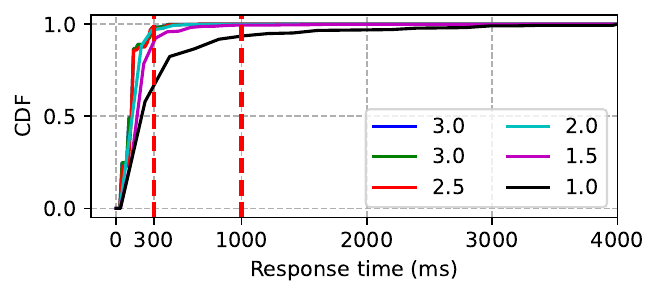}
    \vspace{\evalcaptionvspace}
    \caption{$\Delta E_{00}$ variations}
    \label{fig:delta_e_variation}
    \end{subfigure}%
    \begin{subfigure}{\evalfigurecoeff\textwidth}
        \centering
\includegraphics[width=\evalfigwidthcoeff\linewidth]{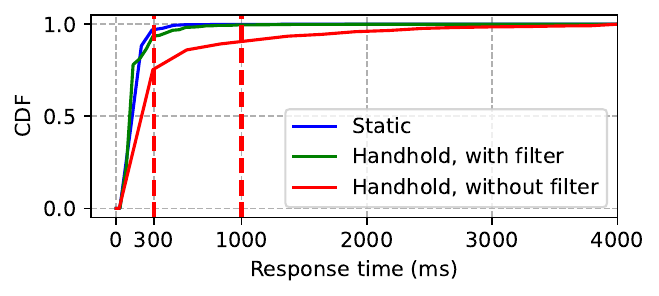}  
\vspace{\evalcaptionvspace}
        \caption{Static vs. handheld}
        \label{fig:static_vs_handheld}
    \end{subfigure}%
    \begin{subfigure}{\evalfigurecoeff\textwidth}
        \centering
        \includegraphics[trim=0cm 0cm 0cm 0cm, width=\evalfigwidthcoeff\columnwidth]{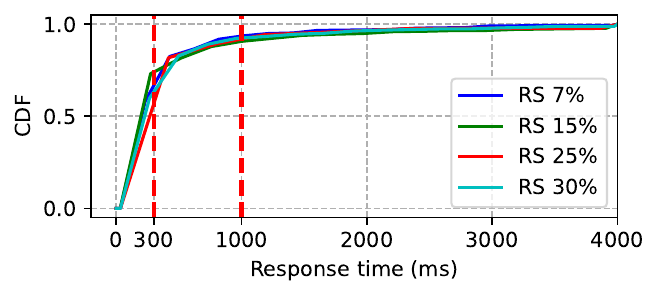}
        \vspace{\evalcaptionvspace}
        \caption{Error correction levels}
        \label{fig:rs_error_levels}
    \end{subfigure}%
    
    \vspace{\totalevalvspace}
    \caption{Response times for different implementation methods. The thresholds of Table \ref{tab:user_thresholds} are marked with red dashed lines.}
    \label{fig:rx_and_setup_pics}
    \vspace{-15pt}
\end{figure*}

\begin{figure}[t]
    \centering
    \begin{subfigure}{0.48\columnwidth}
        \centering
        \includegraphics[width=1\columnwidth]{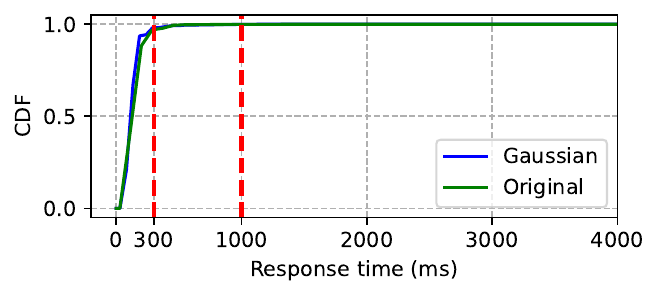}
        \caption{Original vs. Gaussian kernel}
        \label{fig:gaussian_vs_no_gaussian}
    \end{subfigure}
    \begin{subfigure}{0.48\columnwidth}
        \centering
\includegraphics[width=1\columnwidth]{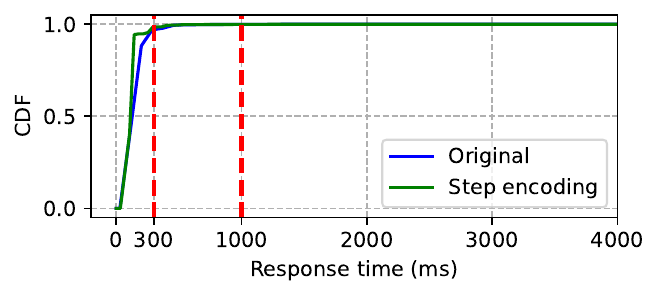}
        \caption{Original vs. step encoding}
        \label{fig:step_vs_no_step}
    \end{subfigure}
    \caption{Response times of different encoding methods.}
    \label{fig:encoding_parameters}
    \vspace{-10pt}
\end{figure}

\begin{figure}[t]
        \centering
        \includegraphics[width=0.5\columnwidth]{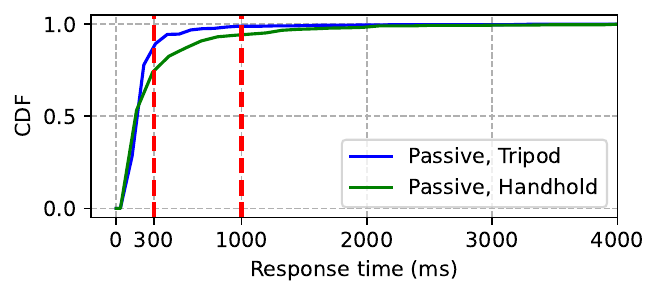}
    \caption{Response time of final system on passive display}
    \label{fig:final_implementation_cdf}
    \vspace{-15pt}
\end{figure}

\subsectionbeforecut
\subsection{Real-time decoding}
\label{sec:real-time_decoding}
In Section \ref{sec:introduction}, we mentioned a sample application for passive screens that synchronizes the audio of a smartphone to a displayed video. In this section, we develop a real-time Android application using the decoding techniques described in Section~\ref{sec:decoding} and benchmark its performance. Unlike the offline decoding used in the prior evaluation, a real-time system must have low response times to keep the user engaged. According to studies in this area\cite{ui_vs_time}, the maximum allowable response times and their interpretations are listed in Table \ref{tab:user_thresholds}.
These thresholds are marked with vertical red dashed lines on all graphs in this section. 
To measure the response times, we calculate the \ac{cdf} of successfully decoding a packet within a given time.
The response time has a positive correlation with reliability (\ac{psr}), i.e. a link of better quality has a shorter decoding time. However, this depends on the phone's processing capabilities and the parameters used in data embedding.
\begin{table}[t]
    \centering
    \begin{tabular}{|c|c|}
    \hline
    Response time & User's perception of the system \\ \hline \hline
      $<$ 300\,ms   & Instantaneous \\ \hline
      300\,ms to 1\,s  & Immediate \\ \hline 
      1 to 5\,s & Transient, not immediate, but the\\ 
        &  user does not disengage from the activity.\\ \hline
    \end{tabular}
    \caption{User's experience thresholds according to \cite{ui_vs_time}}
    \label{tab:user_thresholds}
    \vspace{-10pt}    
\end{table}
The phone we use is a medium-end device (Xiaomi Redmi Note 10 4G) and the Android application is optimized with multithreading to take advantage of the processing power. 

In this section, we run the tests on our modified display to provide an upper bound of response times, since the smart window performs even better, as described earlier. 
The encoding parameters we control and their effect on the system's performance are mentioned below.


\subsubsection{The impact of $\Delta E_{00}$} This value determines how perceptible the encoded data is. We sweep this parameter from a low value of 1 to a large value (most visible) of 3 in steps of 0.5. Then we measure the cumulative distribution of the response times. The results are shown in Fig. \ref{fig:delta_e_variation}.
Interestingly, the \ac{cdf} curves for $\Delta E_{00} \in \{2 , 2.5 , 3\}$ are similar. There is a 95\% chance of successfully decoding packets within 265 ms, 269 ms, and 267 ms, respectively. However, when $\Delta E_{00}$ is decreased to 1.5, the response time increases to 400 ms to have the same 95\% success. Furthermore, for $\Delta E_{00} = 1$, the response time further escalated to 1305 ms.
Although increasing the $\Delta E_{00}$ results in a shorter response time, it also makes the \ac{qr} codes more visible. In Section \ref{sec:survey}, we perform a survey on visibility, where we show that with $\Delta E_{00} = 1.5$, the users \textit{only} perceive a slight improvement compared to $\Delta E_{00} = 2.0$. Therefore, considering that the response time is substantially better in the latter case, we opt for a $\Delta E_{00}$ value of $2.0$ in the rest of the experiments.

\subsubsection{Redundancy of embedded data}
\label{subsec:performance_1_vs_multi_qr}
A single \ac{qr} code occupies a small space and could leave large regions of a frame unused. By duplicating the \ac{qr} code, the unused space can be utilized in favor of redundancy. To investigate this approach, the performance of a single \ac{qr} code is compared with that of a video containing six \ac{qr} codes copied all over the frame. As shown in Figure~\ref{fig:single_qr_vs_multiple}, the \ac{cdf} of a single \ac{qr} code has a probability of 95\% at a response time of 805\,ms, while the link with 6 \ac{qr} codes has the same reliability at 267\,ms. This is a substantial improvement. As a result, we use the version with 6 \ac{qr} codes in the rest of the experiments despite being more noticeable to a user. In our survey, we will show that the slight increase in visibility is a valuable trade-off to achieve a considerably lower response time. 


\begin{figure*}[t]
    \centering
    \begin{subfigure}{\evalfigurecoeff\textwidth}
    \centering
    \includegraphics[width=\evalfigwidthcoeff  \columnwidth]{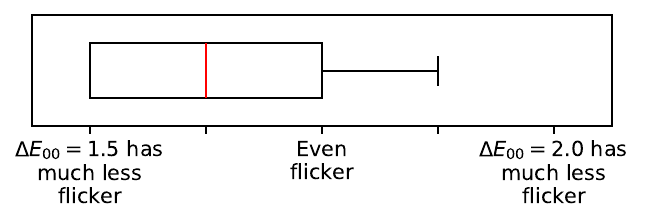}
        \vspace{\evalcaptionvspace}
    \caption{$\Delta E_{00}=1.5$ vs $\Delta E_{00}=2.0$}
    \label{fig:delta_e_flicker}
\end{subfigure}
    \begin{subfigure}{\evalfigurecoeff\textwidth}
        \centering
\includegraphics[width=\evalfigwidthcoeff\columnwidth]{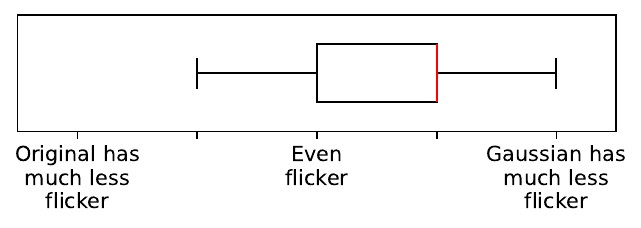}
    \vspace{\evalcaptionvspace}
    \caption{Effect of Gaussian kernel}
    \label{fig:original_vs_gaussian_flicker}
    \end{subfigure}%
    \begin{subfigure}{\evalfigurecoeff\textwidth}
        \centering
        \includegraphics[trim=0cm 0cm 0cm 0cm, width=\evalfigwidthcoeff\columnwidth]{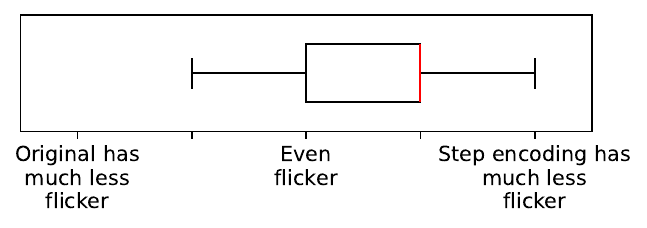}
        \vspace{\evalcaptionvspace}
        \caption{Effect of step encoding}
        \label{fig:step_encoding_flicker}
    \end{subfigure}%
    \begin{subfigure}{\evalfigurecoeff\textwidth}
        \centering
\includegraphics[width=\evalfigwidthcoeff\linewidth]{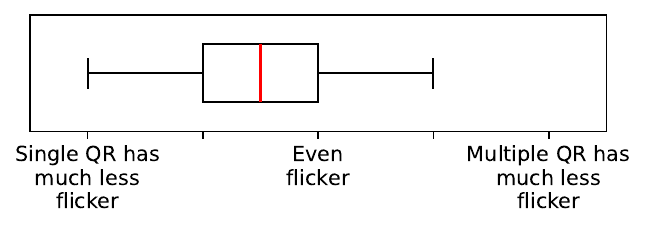}  
\vspace{\evalcaptionvspace}
        \caption{Single QR vs. 6 QRs}
        \label{fig:multi_qr_flicker}
    \end{subfigure}%
    \vspace{\totalevalvspace}
    \caption{Survey results comparing flicker for different parameters and encoding techniques.}
    \label{fig:flicker_compare}
    \vspace{-15pt}
\end{figure*}

\subsubsection{Hand motion filter}
\label{subsec:performance_hand_motion_filter}
To verify the effect of hand motion on the response time, we do three experiments: (i) the phone is placed on a tripod and the filter is not active, (ii) the phone is held by hand and the filter is still not enabled, and (iii) the phone is held by a user and the hand motion filter is enabled. 
The results are shown in Figure~\ref{fig:static_vs_handheld}. In experiment (i), the \ac{cdf} reaches a probability of 95\% at 267\,ms.
In experiment (ii), the same success probability is achieved at 1737\,ms, a substantially longer period. This can be attributed to unaligned frames, causing the phone to struggle to decode. Finally, in experiment (iii), a remarkable improvement is seen, with a \ac{cdf} of 95\% at 372 ms. This represents a significant enhancement compared to the configuration without the hand motion filter and is only slightly inferior to the static setup.

\subsubsection{\ac{qr} error correction level}
\label{subsec:performance_RS}
When investigating the effect of $\Delta E_{00}$, we mentioned that a $\Delta E_{00}=2$ satisfies the real-time requirements but it is not invisible. On the other hand, a $\Delta E_{00}=1$ performs poorly in terms of response time 
while making the codes almost invisible. To investigate this trade-off further, we look into the inherent \ac{rs} error correction of \ac{qr} codes. We hypothesize that by increasing the error correction, the \textit{unrecoverable} codes with a $\Delta E_{00}$ of 1 might become \textit{detectable}, resulting in a lower response time. 

In this regard, we set the $\Delta E_{00}$ to 1, and use four different \ac{rs} levels on the \ac{qr} codes. The \ac{cdf}s of these experiments are shown in Fig. \ref{fig:rs_error_levels}. 
We can see that none of the curves provide good performance according to the real-time classification of Table~\ref{tab:user_thresholds}. For a \ac{cdf} of 85\%, the response times with different error correction levels are: 7\% \ac{rs} error correction yields a response time of 532\,ms, 15\% results in 545 ms, 25\% gives 528 ms, and 30\% results in 530 ms. Contrary to the hypothesis, all error correction levels demonstrate comparable performance. Considering these results, it is better to use a \ac{qr} code with 7\% \ac{rs} error correction as it occupies less space and makes room to place more data.

\subsubsection{Gaussian kernel}
\label{subsec:performance_gaussian_kernel}
Applying a Gaussian kernel to the binary frames makes the encoded data less visible. To verify its effect on response time, we compare two cases with and without a kernel. The results are shown in Fig. \ref{fig:gaussian_vs_no_gaussian}. Without a Gaussian kernel, the \ac{cdf} reaches 95\% at 266\,ms. With the kernel, the response time does not vary and is 267\,ms. Hence, the Gaussian kernel is harmless to the response time while it reduces the visual effects of the encoding.

\subsubsection{Step encoding}
\label{subsec:performance_step_encoding}
The \ac{cdf} of the response times for transmitters with and without step encoding is shown in Figure~\ref{fig:step_vs_no_step}. When step encoding is enabled, a \ac{cdf} of 95\% is reached at 244\,ms. When disabled, 
it gives a response time of 267\,ms. Hence, this enhanced encoding does not affect the response time much and is safe to use. 

\subsection{Final implementation}
\label{sec:final_implementation}
Our final implementation uses the results from the last sections. 
The video embeds 6 \ac{qr} codes with an error correction of 7\%, the $\Delta E_{00}$ is set to 2.0, and a Gaussian filter is applied. 
The real-time pipeline is tested with a 
passive screen and a camera in two conditions: hand-held and on a tripod. These results are shown in Fig. \ref{fig:final_implementation_cdf}. Based on the measured \ac{cdf}, 
there is a 95\% chance of decoding packets successfully within 530\,ms using a tripod and 1071\,ms when held by hand. According to \cite{ui_vs_time}, 
these response times are both classified within or very close to the \textit{immediate} response category

\subsectionbeforecut
\subsection{Survey results}
\label{sec:survey}
Although we propose methods to hide the encoding, visual distortions are inevitable since videos are modified.
To evaluate modifications, we perform a survey where we place two videos side-by-side: one with and the other without the intended modification. Then, we asked a total of 20 users to compare the flicker of the two videos by choosing one out of five levels. 
The results are shown in the box plots of Fig. \ref{fig:flicker_compare}. In these experiments, we consider the following four cases:

\noindent\textbf{a) Apply a $\Delta E_{00}$ of 1.5 or 2?} As mentioned in Section \ref{sec:real-time_decoding}, there is a significant advantage in the response time when we move from a $\Delta E_{00}$ of 1.5 to 2. It is important to assess the user perception as well. As shown in Fig. \ref{fig:delta_e_flicker}, the median of users said there is ``slightly less flicker'' for the lower value of $\Delta E_{00}$. Considering this minor effect, and the big advantage in the response time of $\Delta E_{00}=2$, we opt for this higher value.  

\noindent\textbf{b) Apply Gaussian kernel or not?} Following Fig. \ref{fig:original_vs_gaussian_flicker}, applying a Gaussian kernel helps reduce flicker. 

\noindent\textbf{c) Apply step encoding or not?} Following Fig. \ref{fig:step_encoding_flicker}, users see a better performance when step encoding is applied. 

\noindent\textbf{d) Use single or multiple \ac{qr} codes?} Using more \ac{qr} codes occupies more space in a video, which can cause more flicker as validated by the survey in Figure~\ref{fig:multi_qr_flicker}. However, this flicker is not significant. Considering its lower response time, the redundancy of \ac{qr}s is worth implementing.  


\subsection{\papername in action}
\label{sec:passivecam_in_action}
In Section \ref{sec:introduction}, we mentioned that passive screens are being deployed more and more in public places. This opens opportunities for new applications. Together with members of a company, we propose the development of a real-time link for the following application: when a video is being publicly displayed, it is usually difficult for a user to hear quality audio. Therefore, we use a phone to synchronize the video so users hear the video's sound on their phone. To make this system, we encode synchronization packets--i.e. frame numbers of the video--unobtrusively using the methods discussed before. Then, with our real-time application, the user's phone can read the hidden messages and play the audio accordingly. A demo of this application for a black-and-white and a color video has been put in the links of \cite{video_black_white} and \cite{video_color}. More details of this implementation are described in Appendix~\ref{app:application_implementation}.
\vspace{-5pt}

%% file: chapters/2.Background.tex
\section{Related work}
\label{sec:background}

\textbf{Unobtrusive screen-to-camera communication.} Numerous studies have explored imperceptible data encoding in videos\cite{inframe, inframepp, chromacode, texturecode, deeplight, aircode, uberlight, chromaticity_mod}.
The pioneering works in this area are InFrame \cite{inframe} and its subsequent improvement InFrame++ \cite{inframepp}. They are the first to exploit the flicker fusion property of human eyes with complementary frames. However, TextureCode \cite{texturecode} identified that some flicker was present in those studies. The flicker appeared in smooth areas, hence, TextureCode proposed texture analysis to selectively pick the regions where data should be embedded. The latter improves invisibility but it sacrifices throughput, as it only uses areas with rough texture. To maintain invisibility while leveraging the entire frame, ChromaCode combines luminance with texture analysis \cite{chromacode}. Instead of using a uniform delta value to modulate the entire frame, ChromaCode adjusts the delta based on the expected human perception of color and texture. AirCode \cite{aircode} employs the same video encoding as ChromaCode with the addition of an audio channel transmitting metadata, which compensates for the more error-prone screen channel. Given ChromaCode advancements and the fact that it uses lightness modulation instead of color, we used it as the basis of our work.

Other papers explore color domains to achieve invisibility. Deeplight \cite{deeplight} applies the delta value only to the blue channel as human eyes are less sensitive to blue light. 
Uber-In-Light uses both the blue and red channels to embed data \cite{uberlight}, modifying the green channel just to compensate for the visual distortions. Other studies focus on chromaticity, considering that the maximum perceptible chromatic flicker occurs at approximately 25 Hz \cite{chromaticity_mod}. These color based studies are valuable, however, not applicable to black-and-white smart windows.

Overall, compared to all the above studies, our work the first to design a system for passive screens. Furthermore, our system works in real-time with low-end smartphones.


\textbf{Passive communication with light.} Another area that inspired our work is one using optical devices to modulate ambient light \cite{luxlink,passive_vlc,sunfi,Pixel,RetroVLC,chromalux}. These platforms can work with photodiodes and cameras but all of them focus on point-to-point links. 
To our knowledge, the only paper that moves towards passive screens is Sunbox \cite{sunbox}. This work employs a tiny microdisplay (mm in size) to transmit \ac{qr} codes to a smartphone via ambient light reflections. However, there is no video transmission, only \ac{qr} codes, and the phone has to be placed statically on a holder at a distance of three centimeters from the microdisplay.


%% file: chapters/6.Conclusion.tex
\vspace{-4pt}
\section{Conclusion}
\vspace{-4pt}
In this paper, we explored a novel screen-to-camera communication link by using passive displays. 
Besides, by improving on top of the \ac{soa}, we encoded data unobtrusively in a video, which was then displayed by our passive and active screens. We showed that by using a smart phone, this encoded data could be extracted. As a proof-of-concept, we tested our custom Android application in a real-time audio synchronization scenario. Additionally, we assessed the performance of this system both in real-time and offline, based on several factors and user surveys. Overall, our work is the first to make a step towards new opportunities in \textit{passive} screen-to-camera communication domain, paving the way for pervasive applications such as smart surfaces and in-video advertising.

%% file: chapters/8.Acknowledgement.tex
\section{Acknowledgements} 
This work is part of the \textit{LuxSenz} project, a \textit{TOP-Grant, Module 1, Physical Sciences} with project number 612.001.854, which is financed by the Dutch Research Council (NWO).

%% file: chapters/7.appendix.tex
\subsection{Optimization of texture analysis}

\label{app:optimize_constrast}

In prior literature, regional texture analysis is implemented by constructing a \ac{glcm} for every pixel. \ac{glcm} is a known method that quantifies the \textit{texture} inside an $H\times W$ window around a pixel location (x,y) of the image. H and W are arbitrary numbers denoting the dimensions of the chosen window.   
The resulting \ac{glcm} has a dimension of $N_g \times N_g$, where $N_g$ is the number of distinct grey levels found in the image. Therefore, it is possible to obtain large \ac{glcm} matrices depending on the context of an image. Larger matrices have negative effects on calculations in two directions: (1) They consume time to calculate and access and (2) they take up space in the memory. To optimize for these issues, we simplify the calculations as mentioned below. 

Let us denote an element of the \ac{glcm} matrix by $p_{(x,y)}(i,j)$, where (x,y) is the location of the pixel in the original image and the tuple (i,j) is the index of an element within the \ac{glcm} matrix of that pixel. Following the calculation procedure of \ac{glcm}\cite{chromacode}, each $p_{(x,y)}(i,j)$ can be written as follows:
\begin{equation}
\label{eq:glcm_math}
p_{(x, y)}(i, j) = 
\frac{1}{R_{(x, y)}}
\sum_{r = 0}^{H - 1} \sum_{c = 0}^{W - 2} V(c,r)
\end{equation}
\begin{equation*}
\text{where }
V = \begin{cases}
1, \text{ if } L^*(c, r) = i \text{ and } L^*(c+1, r) = j\\
0, \text{ otherwise}
\end{cases}
\end{equation*}

where $H$ and $W$ are the respective height and width of the arbitrary window around the pixel (x,y), and $R_{(x, y)}$ is the total number of horizontal neighboring in that window. 
In the summation, the index pair (c,r) sweeps over all pixels in the arbitrary window and sets the value of $V$ to 1 only if the pixel at (c,r) has the lightness of i and its right neighbour has the lightness j. In short, the element (i,j) of the \ac{glcm} matrix counts pixel pairs in the arbitrary window such that the left one has the lightness of i and its immediate right neighbour has the lightness of j. 

After this, the contrast equation of the pixel at (x,y) reads as follows, assuming that $H=W=n$:
 \begin{equation}
C(x, y) = 
\sum_{n=0}^{N_g-1} n^2 \{ 
\sum_{i=0}^{N_g}\sum_{j=0}^{N_g} p_{(x,y)}(i, j)
\},\ \ \ \  \text{s.t.} |i-j|=n
\label{eq:contrast}
\end{equation}

Substituting Equation \ref{eq:glcm_math} in Equation \ref{eq:contrast} gives: 

\begin{equation}
C(x, y) = 
\sum_{n=0}^{N_g-1} n^2 \{ 
\sum_{i=0}^{N_g}\sum_{j=0}^{N_g} 
\frac{1}{R_{(x, y)}} 
\sum_{r = 0 }^{H-1}
\sum_{c = 0}^{W-2} V(c,r)
\}
\label{eq:contrast_s1}
\end{equation}
\begin{equation*}
\text{ s.t. } |i-j|=n.    
\end{equation*}
Moving the $\frac{1}{R_{(x, y)}}$ factor out of the summation and rearranging the terms will yield:
\begin{equation}
\label{eq:contrast_s2}
C(x, y) = 
\frac{1}{R_{(x, y)}} 
\sum_{n=0}^{N_g-1} n^2 \{ 
\sum_{r = 0}^{H-1}
\sum_{c = 0}^{W-2} 
\sum_{i=0}^{N_g}\sum_{j=0}^{N_g} V\}
\end{equation}
\begin{equation*}
\text{ s.t. } |i-j|=n.
\end{equation*}
Now the part $\sum_{i=0}^{N_g}\sum_{j=0}^{N_g} V$ will either be 1 if $L^*(c, r) = i \text{ and } L^*(c+1, r) = j$ or 0 otherwise. Therefore the equation in~(\ref{eq:contrast_s2}) can be simplified to:
\begin{equation}
\label{eq:contrast_s3}
C(x, y) = 
\frac{1}{R_{(x, y)}} 
\sum_{r = 0}^{H-1}
\sum_{r = 0}^{W-2} 
\sum_{n=0}^{N_g-1} n^2 
\end{equation}
\begin{equation*}
\text{ s.t. } |L^*(c, r)-L^*(c+1, r)|=n.
\end{equation*}
Substituting for the value of $n$ in Equation \ref{eq:contrast_s3} will result:
\begin{equation}
C(x, y) = 
\frac{1}{R_{(x, y)}} 
\sum_{r = 0}^{H-1}
\sum_{r = 0}^{W-2} 
(L^*(c, r)-L^*(c+1, r))^2
\label{eq:contrast_simplified}
\end{equation}

Using Equation \ref{eq:contrast_simplified}, we can bypass creating a \ac{glcm} for each pixel to 
reduce the computational time significantly.
When benchmarked on a \ac{gpu}, the computational time of the optimized method was 40\,ms for two complementary frames, whereas it was 20190\,ms using the original method of formulating a \ac{glcm} for each pixel, yielding a speed-up of more than 500$\times$.



\subsection{Implementation details of the music application}
\label{app:application_implementation}

In this appendix, we describe the details of our application: a user points his smartphone camera towards the passive screen while a video is playing. The smartphone is then triggered to play the audio synchronized with the video.

In our example we use different songs including the video clip ``Coldplay - The Scientist''. 
The embedded \ac{qr} codes contain a song identifier and its frame's index. This information is represented by a JSON string serving as a lightweight data interchange format.
Each JSON string is structured as \{`s':$s_{id}$,`f':$f_{num}$\}, where the $s_{id}$ is an integer representing the song, and $f_{num}$ is the frame index of the video. Each song is saved on the smartphone prior to the experiment. With the $s_{id}$, the application can select the correct song, and with the frame number, it calculates the current time of the video.


To synchronize the audio with the video, the decoded frame number is converted to an absolute time value in milliseconds ($t_{abs}$) using the following equation:
\begin{equation}
t_{abs} = \frac{2 f_{num}}{FPS_{TX}} \cdot 1000
\label{eq:frame_num_to_abs}
\end{equation}
where $FPS_{TX}$ represents the screen's frame rate. The factor of 2 arises as each \ac{qr} code is transmitted using two complementary frames. By multiplying the resulting value by 1000 the absolute time ($t_{abs}$) is found in milliseconds. In the final step, the application must compensate for the time taken to decode. This elapsed time is measured using the internal system clock of the Android phone and is then added to $t_{abs}$ to have an estimate of the current timestamp of the video playback. After this, the phone starts playing the audio from the calculated time in the last step. Thus, the synchronization is achieved.    
